\documentstyle[anta, twoside]{article}

\input{psfig.sty}

\def\Journal#1#2#3#4{(#1) {#2} {\bf #3}, #4}

\def\AAp{\em Astron. Astrophys.}

\def\ApJ{\em Astrophys.~J.}

\def\ARAaAp{\em Annnu. Rev. Astron. Astrophys.}

\def\MNRAS{\em Mon. Not. R.~Astron. Soc.}
\def\Nat{\em Nature\/}
\def\PASP{\em Publ. Astron. Soc. Pac.}
\def\PASJ{\em Publ. Astron. Soc. Japan}
\def\qjras{\em Qua.J. R.~Astron. Soc.}

\newcommand{\HII}{{\rm H\,\scriptstyle II}}
\begin{document}

\markboth{J.L. Han}{The Large-Scale Magnetic Field Structure of Our Galaxy}

\thispagestyle{plain}
\setcounter{page}{3}

\title{The Large-Scale Magnetic Field Structure of Our Galaxy: \\
 Efficiently Deduced from Pulsar Rotation Measures}

\author{J.L. Han}

\address{National Astronomical Observatories, Chinese Academy of Sciences\\
Jia 20 Da-Tun Road, Beijing 100012, China\\ }

\maketitle

\abstract{In this review, I will first introduce possible methods to probe
the large-scale magnetic fields in our Galaxy and discuss their
limitations. The magnetic fields in the Galactic halo, mainly revealed by
the sky distribution of rotation measures of extragalactic radio sources,
probably have a global structure of a twisted dipole field. The large-scale
field structure in the Galactic disk has been most efficiently deduced from
pulsar rotation measures (RMs). There has been a lot of progress since the
1980s when the magnetic field in the local area of our Galaxy was first
traced by a very small sample of local pulsars. Now we have pulsars
distributed in about one third of the whole Galactic disk of the
interstellar medium, which shows that the large-scale magnetic fields
go along the spiral arms and that the field directions reverse from arm
to arm. The RMs of newly discovered pulsars in the very inner Galaxy
have been used to show the coherent magnetic field in the Norma arm.
The magnetic fields in the Galactic disk most probably have a
bisymmetric spiral structure.}

\section{Introduction}

The origin of magnetic fields in the universe is a long-standing problem.
It is clear today that magnetic fields play a crucial role in the evolution
of molecular clouds and star formation (e.g. Rees 1987).  The diffuse
magnetic fields are the physical means to confine the cosmic rays
(e.g. Strong et al. 2000).  The magnetic fields in galactic disks also have
a significant contribution to the hydrostatic balance in the interstellar
medium (Boulares \& Cox 1990). However, it is not clear whether magnetic
fields exist in the very early universe, e.g. the recombination phase,
and whether the fields affect the structure formation and galaxy
formation afterward.

To understand the magnetic fields, the first step is to correctly and
properly describe their properties based on reliable observations. Magnetic
fields of galactic scales ($\sim 10$~kpc) are the most important connection
between the magnetic fields at cosmological scales and the fields in
currently observable objects. In the last two decades, there have been many
observations on magnetic fields in galaxies (see references in reviews by
Beck et al. 1996; Han \& Wielebinski 2002). Theoretically the magnetic
fields in galaxies are believed to be re-generated and maintained by dynamo
actions in the interstellar medium (e.g. Ruzmaikin et al. 1988; Kulsrud
1999). The helical turbulence (the $\alpha$-effect) and differential
rotation (the $\Omega$-effect) are two key ingredients for dynamos in
galaxies (e.g. Krause \& R\"adler 1980).

Our Galaxy is the unique case for detailed studies of magnetic fields, as I
will show in this review.

It is certainly desirable to know the structure of our Galaxy and to compare
it with the magnetic structures. However, as we live near the edge of the
disk of the Milky Way, it is impossible for us to get a clear bird-view of
the global structure of the whole Galaxy. As has been shown by various
tracers, our Galaxy obviously has a few spiral arms with a pitch angle of
about 10$^\circ$. But there is no consensus on the number of spiral arms in
our Galaxy and whether and how these arms are connected in the opposite
side.  Apart from the thin disk, there is a thick disk or halo, filled by
low-density gas and possibly weak magnetic fields.

Pulsars are the best probes for the large-scale magnetic fields in our
Galaxy.  At present, many new pulsars have been discovered up to distances
further than the Galactic center (e.g. Manchester et al. 2001; Morris et
al. 2002; Kramer et al.  2003), which can be used to probe the large-scale
magnetic fields in about half of the Galactic disk. The magnetic field may
further give us some hints to the global structure of our Galaxy. Hundreds
of pulsars discovered at high Galactic latitudes (e.g. Edwards et al. 2001)
can be used to study the magnetic fields in the Galactic halo.

\subsection{Comments on definitions of useful terms}

Before we start to discuss observational results of magnetic fields, it
is worth to clarify the definition of some useful terms.

\medskip
{\noindent \bf Large scale vs. small scale:}\\ How large is the ``{\it
large scale}''?  Obviously it should be a scale, relatively much larger than
some kind of standard.  For example, the large-scale magnetic field of the
Sun refers to the global-scale field or the field with a scale-length
comparable to the size of Sun, up to $10^9$~m, rather than small-scale
magnetic fields in the solar surface.  For the magnetic fields of our
Galaxy, we should define the {\it large scale} as being a {\sl scale
larger than the separation between spiral arms}. That is to say,
large scale means a scale larger than 2 or 3 kpc.

Note that in the literature, {\it large scale} is sometimes used for
large {\it angular} scale when discussing the structures or prominent
features {\it in the sky plane}, e.g., the large-scale features in radio
continuum radio surveys. These {\it large angular-scale} features are
often very localized phenomena, and not very large in linear scale.

\medskip
{\noindent \bf Ordered vs. random fields:} \\ Whether a magnetic
field is ordered or random depends on the scales concerned. A
uniform field at a 1-kpc scale could be part of random fields at a
10-kpc scale, while it is of a very large scale relative to the
pc-scale magnetic fields in molecular clouds.

{\it Uniform fields} are {\it ordered fields}. {\it Regularly ordered
fields} can coherently change their {\it directions}, so they may not be
{\it uniform fields}. Deviations from {\it regular fields} or {\it orderd
fields} are taken as {\it random fields}. The fluctuations of fields
at scales 10 times smaller than a concerned scale are oftern taken as
{\it random fields}.

Note also that in the literature the measurements of so-called {\it
polarization vectors} only give the {\it orientations} rather than {\it
directions} of magnetic fields, so they are not real vectors.

\medskip
{\noindent \bf Azimuthal and toroidal fields:}\\ Often the magnetic fields
in our Galaxy are expressed in cylindrical coordinates ($\theta,\ r,\
z$). The {\it azimuthal component}, $B_\theta$, of magnetic fields dominates
in the Galactic disk, where the {\it radial and vertical components}, $B_r$
and $B_z$, are generally weak. The {\it toroidal component} refers to
the structures (without $B_z$ components) confined to a plane parallel
to the Galactic plane, while the {\it poloidal component} refers to the
axisymmetrical field structure around $z$, such as dipole fields
(without $B_\theta$ component).

Concerning measurements relative to the line of sight, only one field
component, either {\it perpendicular} or {\it parallel} to the line of
sight, can be detected by one method (see below).  \\

The above terms are artificially designed for convenience when studying
magnetic fields. Real magnetic fields would be all connected in space,
with all components everywhere.

\subsection{Observational tracers of magnetic fields}

{\noindent \bf Zeeman splitting:}\\ It measures the {\it parallel
component} of magnetic fields in an emission or absorption region
by using the the splitting of spectral lines.

Up to now, measurements of magnetic fields {\it in situ} of masers
(e.g. Fish et al. 2003) and molecular clouds (e.g. Bourke et al. 2001)
are available. The relationship between the field strength and gas
density has been supported by observational data (e.g. Crutcher 1999).
Despite strong efforts it failed to relate the magnetic fields in situ
and the large-scale fields (see Fish et al. 2003), as suggested by
Davies (1974) and later promoted by Reid \& Silverstain (1990).

\medskip
{\noindent \bf Polarization at infrared, sub-mm and mm wavebands:}\\ Dust
particles are preferentially orientated due to the ambient magnetic
fields. The thermal emission of dust then naturally has linear
polarization. Infrared, mm, submm instruments are the best to detect
thermal emission. Polarization shows directly magnetic fields projected in
the sky plane. Due to the short wavelengths, such a polarized emission does
not suffer from any Faraday rotation when passing through the interstellar
space.

The recent advance in technology has made it possible to make direct
polarization mapping at infrared, sub-mm and mm wavebands (e.g. Hildebrand
et al. 1998; Novak et al. 2003). At present, measurements can only be made
for bright objects, mostly of molecular clouds. But in future it could be
more sensitive and powerful to measure even nearby galaxies. A
combination of polarization mapping for the perpendicular component of
magnetic fields with the parallel components measured from Zeeman
splitting will give a 3-D information of magnetic fields in molecular
clouds (or galaxies in future), which is certainly crucial to study the role
of magnetic fields in the star-formation process.

The available measurement, which is really related to large-scale magnetic
fields, is the polarization mapping of the central molecular zone by Novak
et al. (2003). The results revealed possible toroidal fields parallel to
the Galactic disk. This field is probably part of an A0 dynamo field in the
Galactic halo, complimented to the poloidal fields traced by vertical
filaments in the Galactic center (e.g. Sofue et al. 1987; Yusef-Zadeh \&
Morris 1987).

\medskip
{\noindent \bf Polarization of starlight:}\\ The starlight is scattered by
interstellar dust when traveling from a star to the earth. The dust particles
are preferentially orientated along the interstellar magnetic fields, which
induce the polarization of the scattered star light: more scattering, more
polarization.

Starlight polarization was the start for the studies of the large-scale
magnetic field. Apparently, the starlight can trace prominent magnetic
features on large angular scales, over all the sky! Directly from the
measurements in the Galactic pole regions, we can easily see the direction
of the local magnetic fields in our Galaxy.  However, because the measured
stars are mostly within 1 or 2~kpc from the Sun, it is not possible to
trace magnetic fields further away. Polarization measurements of stars
near the Galactic plane give us only the information that the magnetic
fields in the Galactic disk are mainly orientated parallel to the
Galactic plane.

\medskip
{\noindent \bf Synchrotron radiation:}\\ Synchrotron radiation from
relativistic electrons shows the {\it orientation} of magnetic fields in the
emission region.  Assuming energy equipartition, one may estimate the field
strength from the emission flux. Furthermore, from polarized intensity, the
energy in the ``uniform field'' together with that of anisotropic random
fields can be estimated.

Many nearby galaxies have been observed in polarized radio emission.  The
emission suffers from Faraday rotatation within the medium of the galaxy and
from the Milky Way. After correcting the foreground RMs, one can obtain a
map of intrinsic {\it orientations} of the {\it transverse component} of
magnetic fields, though it is often called the ``vector'' map of the
magnetic field. The anisotropic random magnetic fields, such as compressed
random fields by large-scale density waves, can also produce such
an observed polarization map. So, with a polarization map with a coherent
``vector'' pattern, one cannot claim the large-scale magnetic field.
However, a map of the RM distribution with a regular pattern, especially for
the inclined galaxies, provides strong evidence for the large-scale magnetic
fields in the halo or the thick disk of a galaxy (see Krause, this volume).

As we will show below, the pulsar RM distribution indeed shows
large-scale magnetic fields in our Galaxy, with coherent field {\it
directions} going along the spiral arms. This indirectly proves that the
polarization maps of nearby galaxies can be at least partially due to the
large-scale magnetic field. The strength of ``regular fields'' calculated
from the polarization percentage may be overestimated, however, as one cannot
quantify the contributions from anisotropic random magnetic fields (see Beck
et al. 2003).

\medskip
{\noindent \bf Faraday rotation of polarized sources:}\\ Faraday rotation
occurs when radio waves travel through the magnetized medium.
The RM, which is measured as being the rate of polarization-angle change
against the square of wavelength, is an integration of magnetic field
strength together with electron density along the line of sight from the
source to the observer, i.e., $RM = a \int_{\rm source}^{\rm observer}
\; n_e \; B_{||} \; dl. $ Here $a$ is a constant, $n_e$ the electron
density, $dl$ is a unit length of the line of sight. Obviously, RMs
measure the average diffuse magnetic field if the electron density is known.

In the following I will concentrate on how the RMs of pulsars and extragalactic
radio sources can be used to probe the large-scale magnetic fields in our
Galaxy. Considering various difficulties in other methods, we are really
lucky that we can use the RMs of an increasing number of pulsars to probe the
large-scale magnetic field in the Galactic disk. It is very hard to conduct
observations of Zeeman splitting, and afterwards it is not possible to
relate these measurements to the large-scale magnetic fields. Starlight
polarization does measure the diffuse magnetic field, but it is not
possible to give any other information than the averaged orientation of the
field in just 1 or 2 kpc. Diffuse radio emission from our Galaxy can only
show the polarized emission from local regions on large {\it
angular} scales, but not on large linear scales in the Galactic disk.

\section{The magnetic field in the Galactic halo: based on the RM sky
  distribution}

In the sky the Milky Way is the largest edge-on Galaxy. This gives
us the unique chance to study the magnetic fields in a galactic
halo in detail, which is not possible at all for nearby extragalaxies.
There is a huge number of extragalactic radio sources as well as
hundreds of pulsars, which can be potentially used to probe the
magnetic fields in the Galactic halo and in the Galactic disk.
\begin{figure}[bt]
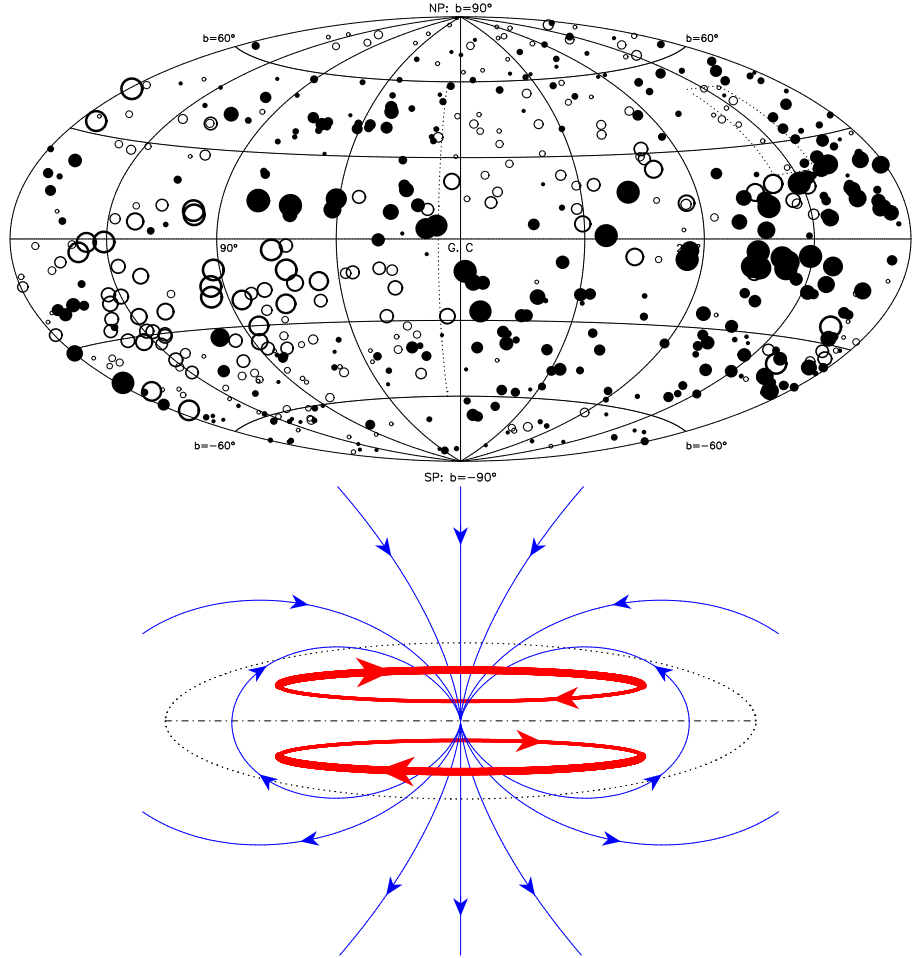

\centerline{\psfig{figure=JLHan_fig1a.ps,angle=270,width=12truecm}}
\centerline{\psfig{figure=JLHan_fig1b.ps,angle=270,width=8.5truecm}}
\caption{The sky distribution of RMs of extragalactic radio sources
and the magnetic field model in the Galactic halo, as discussed
by Han et al. (1997).}
\end{figure}

The RM of an extragalactic radio source consists of a RM contribution
intrinsic to the source, the RM from the intergalactic space from the
source to the Galaxy, and the RM within the Galaxy. The first term
should be random and hence reasonably small on average, because a
source can be randomly orientated in space with any possible field
configuration. We observe a random quantity. The second term is
ignorable on average. Intergalactic magnetic fields are too weak
to be detectable now. A source can be at any possible location in the
universe. Even if there is a weak field in the intergalactic space, an
integration over the path-length of the intergalactic magnetic fields
with random directions together with extremely thin gas should give a
quite small combination. Therefore, the common contribution to
RMs of extragalactic radio sources is from our Galaxy. So, the {\it
averaged} sky distribution of RMs of extragalactic sources (see Fig.~1)
should be the best presentation for the Galactic magnetic field in the
Galactic halo.

Though there are many distinguished characteristics related to specific
regions in the RM sky, we noticed that the most prominent feature
is the antisymmetry in the inner Galactic quadrants (i.e.
$|l|<90^{\circ}$). The positive RMs in the regions of
($0^{\circ}<l<90^{\circ},\; b>0^{\circ}$) and
($270^{\circ}<l<360^{\circ}, \; b<0^{\circ}$) indicate that the
magnetic fields point towards us, while the negative RMs in the
regions of ($0<l<90^{\circ}, \; b<0^{\circ}$) and ($270<l<360^{\circ},
\; b>0^{\circ}$) indicate that the magnetic fields point away
from us. Such a high symmetry to the Galactic plane as the Galactic
meridian through the Galactic center cannot simply be caused by
localized features as previously thought. The antisymmetric pattern is
very consistent with the magnetic field configuration of an A0 dynamo,
which provides such toroidal fields with reversed directions above and
below the Galactic plane (Fig.~1). The toroidal fields possibly extend
to the inner Galaxy, even towards the central molecular zone (Novak et
al. 2003).

This magnetic field model is also supported by the nonthermal radio
filaments observed in the Galactic center region for a long time, which
have been thought to be indications for the poloidal field in dipole
form (Yusef-Zadeh \& Morris 1987; Sofue et al. 1987).

We noticed that the antisymmetric RM sky is also shown by pulsar RMs at high
Galactic latitudes ($|b|>8^{\circ}$). This implies that the magnetic fields
responsible for the antisymmetry pattern could be nearer than the pulsars.
Indeed, the fields nearer than the pulsars contribute to the RMs, but
this is not the only contribution. If it is only a local effect, then
there would be no symmetric RM distribution beyond the pulsars. We made
computer simulations, which show that large-scale magnetic field structures
further away than the pulsars (i.e. the more inner Galaxy) should
result in a strong antisymmetry of RM sky within 50$^{\circ}$ from the
Galactic center, and that the magnitudes of RMs are systematically
increasing towards the Galactic plane and the Galactic center.

\begin{figure}[bt]
\centerline{\psfig{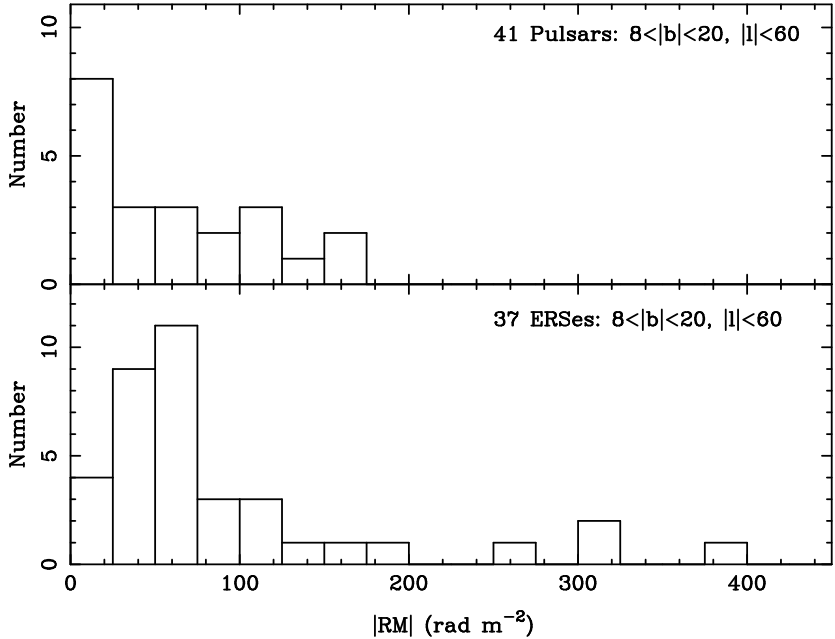}}  
\caption{The RM amplitudes of extragalactic radio sources in the very inner
Galaxy are systematically larger than those of pulsars, indicating that the
antisymmetric fields extended towards the Galactic center, far beyond the
pulsars.}
\end{figure}

To judge if antisymmetry is produced by a large-scale magnetic field, it
would be necessary to subtract the foreground pulsar RMs induced by
local magnetic fields from the RMs of extragalactic radio sources and
then check the antisymmetry of the residual RM map. If there is
antisymmetry in the residual RM map, then it is large-scale, otherwise
it is local. However, both the RM data of pulsars and extragalactic
radio sources are so sparse that such a subtraction cannot give a clear
result at the moment. We checked the magnitude distribution of RMs of
extragalactic radio sources, which is indeed systematically larger than
those of pulsars, as expected from the large-scale halo field (Fig.2).

More RM data of both pulsars and extragalactic radio sources towards medium
Galactic latitudes in the inner Galaxy are desired to check this crucial
issue of Galactic dynamo.

If such a large-scale magnetic field model is confirmed by more data, then
our Galaxy is the first galaxy in which the dynamo signature is clearly
identified. In other words, it is the first time to identify a dynamo at
a galactic scale. This is very difficult for other galaxies. We studied all
possible RM data for M31, and got some evidence for the magnetic field
configuration of a possible S0 dynamo (Han et al. 1998).

\section{Magnetic fields in the Galactic disk: based on pulsar RMs}

Pulsars are the best probes for Galactic magnetic fields. First of all,
pulsars are highly polarized in general. So their RMs are relatively easy
to measure, in contrast to the great difficulties to do Zeeman splitting
measurements. Second, pulsars do not have any intrinsic RM.  So, what one
gets from RMs is just the contribution from the interstellar medium, an
integration of diffuse magnetic fields along the path from a pulsar to the
observer, rather than {\it in situ} measurements in emission regions. Third,
the total amount of the electron density between a pulsar and an observer
can be measured independently by the pulsar dispersion measure,
$DM=\int_{\rm psr}^{\rm obs}\; n_e \; dl$. This leads to a direct
measure of the averaged magnetic field along the line of sight by using
$\langle B_{||} \rangle = 1.232 RM/DM$. There have been many pulsars
discovered, which are widely spread in our Galaxy. After measuring
their RMs, it will not be very hard to find a 3-D magnetic field
structure in our Galaxy.

\subsection{Historical landmarks of using the pulsar RMs for the
large-scale magnetic fields}

A close look at the historical landmarks of using pulsar RMs to study the
Galactic magnetic fields will show the progress in the last two or three
decades.

Soon after the pulsars were discovered, Lyne \& Smith (1969) detected their
linear polarization. They noted that this ``opens up the possibility of
measuring the Faraday rotation in the interstellar medium'' which ``gives a
very direct measure of the interstellar magnetic field'', because
$\int_{\rm obs}^{\rm obj} n_e dl $ can be measured by $DM$ so that $\langle
B_{||}\rangle$ can be directly obtained from the ratio $RM/DM$. Manchester
(1972, 1974) first systematically measured a number of pulsar RMs for
Galactic magnetic fields and concluded that the local field (within 2 kpc!)
is directed toward about $l\sim90\degr$. Thomson \& Nelson (1980) modeled
the pulsar RMs mostly within 2 kpc and found the {\it first} field
reversal near the Carina-Sagittarius arm.  The largest pulsar RM dataset was
published by Hamilton \& Lyne (1987), mostly for pulsars at about 5 kpc
and some up to 10~kpc. Then Lyne \& Smith (1989) used pulsar RMs to
further study the Galactic magnetic field. They confirmed the first
field reversal in the inner Galaxy and found evidence for the field
reversal in the outer Galaxy by a comparison of pulsar RMs with those
of extragalactic radio sources. Rand \& Kulkarni (1989) analyzed 185
pulsar RM data and proposed the ring model for the Galactic magnetic
field. Rand \& Lyne (1994) observed more RMs of distant pulsars and
found evidence for the clock-wise field near the Crux-Scutum arm (at
about 5~kpc). Han \& Qiao (1994) and Indrani \& Deshpande (1998)
reanalyzed the pulsar RM data and found that the RM data are more
consistent with the bisymmetric spiral model than with the ring model.

Han et al. (1997) first noticed that the RM distribution of high-latitude
pulsars is dominated by the azimuthal field in the halo. Afterwards, any
analysis of pulsar RMs for the disk field was limited to pulsars at lower
Galactic latitudes ($|b|<8\degr$). Han et al. (1999) then observed 63 pulsar
RMs and divided all known pulsar RMs into those lying within higher and
lower latitude ranges for studies of the halo and disk field, respectively,
and they confirmed the bisymmetric field structure and refined estimates of
the vertical field component.

\subsection{The large-scale magnetic field models}

\begin{figure}[tb]
\centerline{\psfig{figure=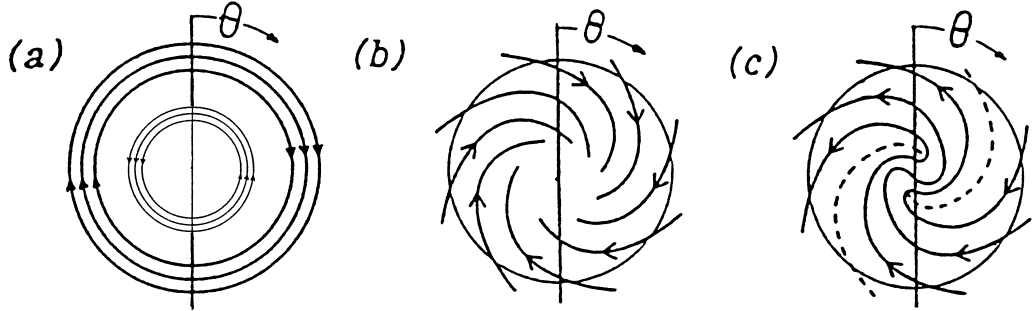,width=14truecm}}
\caption{A sketch of three models for Galactic magnetic fields,
namely, (a) the concentric ring model, (b) the axisymmetric spiral
model, and (c) the bisymmetric spiral model.}
\end{figure}

There have been three models to describe the global magnetic field structure
of our Galaxy. In the early stage, Simard-Normandin \& Kronberg (1980) showed
that the RMs of extragalactic radio sources and pulsars are consistent with
the bisymmetric spiral model. This was later confirmed by Sofue \&
Fujimoto (1983). The currently available pulsar RM data are mostly consistent
with a bisymmetric spiral model as we will discuss below.
While Vall\'ee (1991, 1995) has argued for an axisymmetric spiral
model. In this model, the field reversal occurs only in the range of Galactic
radii from 5 to 8 kpc (Vall\'ee 1996). No field reversals are allowed beyond
8 kpc or within 5 kpc from the Galactic Center. This is in contrast to the
field reversals suggested beyond the solar circle and detected interior to
the Crux-Scutum arm (e.g. Han et al. 1999, 2002). We noticed that recent
arguments favour no field reversal outside the Perseus arm (e.g. Brown et
al. 2003), which need more RM data of pulsars in the Perseus arm to check.
Extragalactic radio source data are not enough to make a solid conclusion.
The concentric ring model, proposed by Rand \& Kulkarni (1989) and Rand \&
Lyne (1994), has a pitch angle of zero, but the observed pitch angle of
fields of $-8^{\circ}$ favours a spiral form of the field structure. Both
the ring model and axisymmetric spiral model show that the magnetic
field lines go across the spiral arms, which seems not to be physically
possible.

\begin{figure}[thb]
\centerline{\psfig{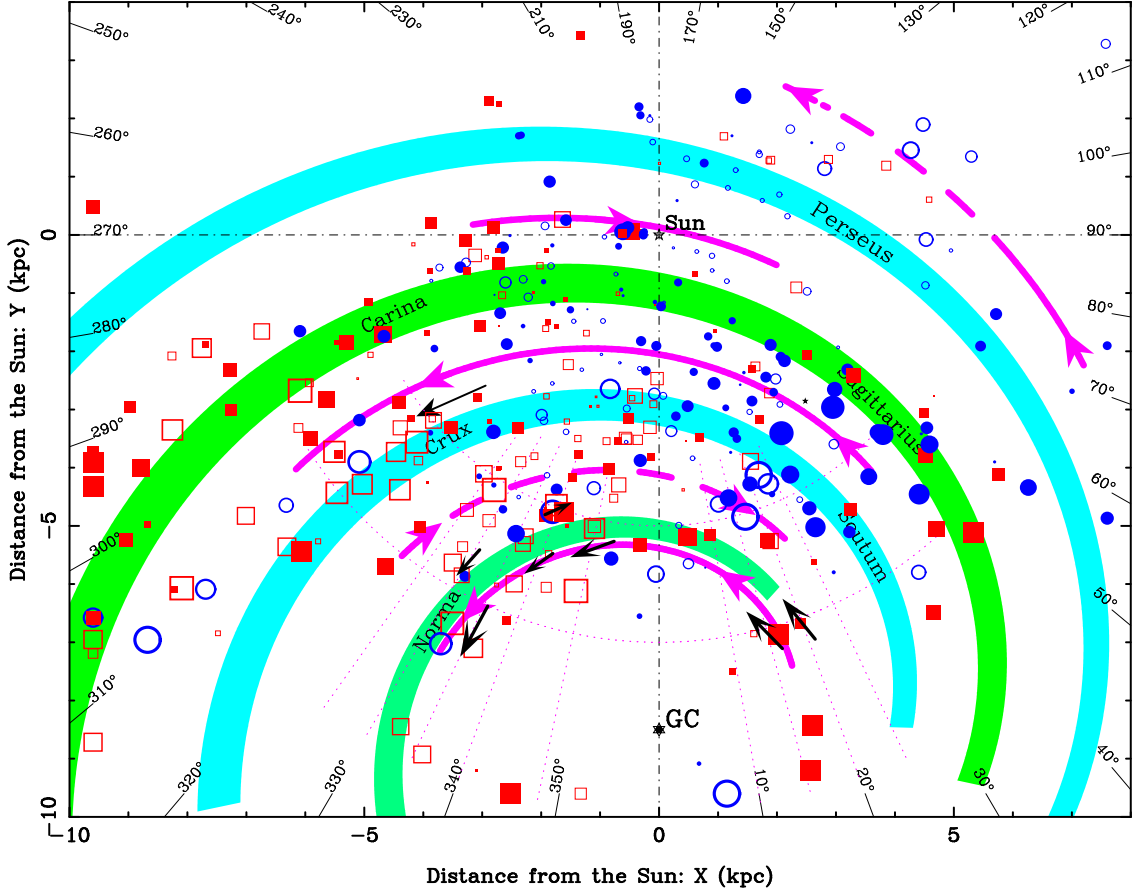}}  
\caption{The RM distribution of pulsars projected onto the Galactic plane.
Red data (squares) are newly observed, and blue (circles) are previously
published. Filled symbols stand for positive RMs and open ones for
negative RMs.  The large-scale magnetic fields are drawn by arrows, which was
inferred from RM data. Solid-line arrows stand for confirmed field
structures, while dashed-line arrows stand for proposed field
structures in controversy and to be confirmed. The pulsar distances
were estimated by a new electron density model (NE2001: Cordes \& Lazio
2002). The magnetic fields are very probably going along the spiral
arms, with {\it coherent directions} over more than 10~kpc interior to
Carina-Sagittarius arm.}
\end{figure}

\subsection{Current status and future directions}

Up to now, among about $\sim$1450 known pulsars, 535 pulsars have measured
values of RM and 373 of them are located at lower latitudes
($|b|<8\degr$). This includes 200 RM data from Parkes observations, which
will be published soon (Han et al. in prep.).  Significant progress has been
made in the last decade on the magnetic fields in the Galactic disk, mainly
because many pulsars have been discovered in the nearby half of the whole
Galactic disk (e.g. Manchester et al. 1996; Lyne et al. 1998; Manchester et
al. 2001) and extensive observations of pulsar RMs (e.g. Hamilton \& Lyne
1987; Rand \& Lyne 1994; Han et al. 1999) were conducted.

Analysis of pulsar RMs needs to consider three important factors for
the diagnosis of the large-scale field structure. First, one normally
assumes that the azimuthal field component $B_{\phi}$ is greater than
the vertical and radial components $B_z$ or $B_r$.  This is reasonable
and has been justified (Han \& Qiao 1994; Han et al.  1999). Second, it
is {\it the gradient of the average or general tendency of RM
variations} versus pulsar DMs that traces the large-scale field.  The
scatter of the data about this general tendency is probably mostly due
to the effect of smaller scale interstellar structure. Finally, the
large-scale field structure should produce a coherence in the gradients
for many independent lines of sight (see e.g. $l=\pm20\degr$ near the
Norma arm in Fig.~4).

From the most updated RM distribution (see Fig.~4), we can conclude that
magnetic fields between the Perseus arm and Carina-Sagittarius arm have a
clock-wise direction when looking from the Northern galactic pole.
Apparently this at least holds for about 5~kpc along the spiral
arms. Between the Carina-Sagittarius arm and the Crux-Scutum arm, the
positive RMs near $l \sim 50^{\circ}$ and negative RMs near $l\sim
315^{\circ}$ show the coherently counter-clockwise magnetic field along the
spiral arm over more than 10~kpc! From the RMs of pulsars discovered by the
Parkes multibeam survey, the counter-clockwise magnetic field along the
Norma arm (i.e. the 3-kpc arm) has been clearly identified (Han et
al. 2002). There have been some indications for clockwise magnetic fields
between the Crux-Scutum arm and the Norma arm, while more RM data are
obviously desired for a definite conclusion.

In the outer Galaxy the magnetic fields directions in or outside the
Perseus arm have been in controversy recently. The magnetic field
reversals suggested by Lyne \& Smith (1989) have been confirmed by Han
et al. (1999) and Weisberg et al. (2004) using available pulsar RMs
mostly near $l\sim70^{\circ}$. While Mitra et al. (2003) and Brown et
al. (2003) have argued for no reversal near or outside the Perseus arm
from the RM data of pulsars and extragalactic radio sources in the
region of $145^{\circ}<l<105^{\circ}$. The average of RM values seems
not to be significantly different for the foreground pulsar RMs near
the Perseus arm and to the background extragalactic radio sources.
This fact probably indicates two field reversals outside the Perseus
arm which cancels their RM contributions. It is necesary to compare RM
data of pulsars in the Perseus arm and background extragalactic radio
sources between $45^{\circ}<l<110^{\circ}$ for that purpose. A solid
conclusion about the magnetic field configurations in this region would
come out soon after many more pulsars in this region will be
discovered in a future Arecibo L-band multibeam pulsar survey.

\subsection{Discussions}

Beside the large-scale magnetic field, naturally there are small-scale
magnetic fields in our Galaxy. The strength of the large-scale magnetic
field has been estimated to be $1.8\pm0.5\;\mu$G (Han \& Qiao 1994; Indrani
\& Deshpande 1998), while the total field strength estimated from
cosmic-rays or using the equipartition assumptions is about 6$\mu$G.  We
have composed the energy spectrum of Galactic magnetic fields at different
scales, from 0.5~kpc to 15~kpc (Han et al. 2003). Based on this
spectrum, we estimate the fluctuations of magnetic fields have an rms
field strength about 6~$\mu$G, which is very consistent with estimates
for the total field strength by other methods. This confirms that the
magnetic field strengths estimated from pulsar rotation measures are
statistically fine for the diffuse interstellar medium. The field
strength of regular magnetic fields estimated from the percentage of
polarized continuum emission of nearby galaxies then probably has been
two or three times overestimated.

Recently, Mitra et al. (2003) have shown that two or three pulsar RMs are
affected by $\HII$ regions as these pulsars can be easily identified by
their large DMs. In fact, the large DM should lead to an overestimated
distance for the pulsar in a given electron density model. However,
only a very small number of pulsars can be affected by chance,
according to simulations made by Cordes \& Lazio (2002).

The coherent variation of RMs versus DMs in different directions not only
provides the information about the large-scale magnetic fields, but also
indicates that the data scattering from the general tendency of
variation due to small-scale regions does not influence the analysis of
RMs for the large-scale magnetic fields.

\section{Conclusions}

Pulsars provide unique probes for the {\it large-scale} interstellar
magnetic field in the Galactic disk. Other methods seem to have many
difficulties for that purpose. The increasing number of RMs, especially of
newly discovered distant pulsars, enables us for the first time to explore
the magnetic field in nearly one third of the Galactic disk. The fields are
found to be {\it coherent in directions} over a linear scale of more than
$\sim 10$ kpc between the Carina-Sagittarius and Crux-Scutum arms from
$l\sim45\degr$ to $l\sim305\degr$ and more than 5~kpc along the Norma
arm. The magnetic fields reverse their directions from arm to arm. The
coherent spiral structures and field direction reversals, including the
newly determined counter-clockwise field near the Norma arm, are
consistent with a bisymmetric spiral model for the disk field.

At high latitudes, the antisymmetric RM sky is most probably produced by the
toroidal field in the Galactic halo. Together with the dipole field in the
Galactic center, it strongly suggests that an A0 dynamo is operating in
the halo of our Galaxy.

\section*{Acknowledgments}
I am very grateful to many colleagues, especially, Prof. R.N. Manchester
and Prof. G.J. Qiao for working together with me to improve the knowledge
on the magnetic fields of our Galaxy, some of which was presented here.
The author is supported by the National Natural Science Foundation of China
(10025313) and the National Key Basic Research Science Foundation of China
(G19990754) as well as from the partner group of MPIfR at NAOC. I wish to
thank the organizers of the conference to invite me to present this review,
and many thanks to Dr. Wolfgang Reich and Ms. Gabi Breuer for carefully
reading the manuscript.

\section*{References}\noindent

\references

Beck R.,  Brandenburg A.,  Moss D.,  Shukurov A.,    Sokoloff D.
  \Journal{1996}{\ARAaAp}{34}{155}.

Beck R., Shukurov A., Sokoloff D., Wielebinski R. \Journal{2003}{\AAp}{411}{99}.

Boulares A., Cox D.P. \Journal{1990}{\ApJ}{365}{544}.

Bourke T.L.,  Myers P.C., Robinson G., Hyland A. R. \Journal{2001}{\ApJ}{552}{916}.

Brown J. C., Taylor A. R., Wielebinski R., Mueller P.
\Journal{2003}{\ApJ}{592}{L29}.

Cordes J.M., Lazio T.J.W. (2002) {\ApJ}, submitted.

Crutcher R.M. \Journal{1999}{\ApJ}{520}{706}.

Davies R.D. (1974) {\it IAU Simp.} {\bf 60}, 275.

Edwards R.T., Bailes M., van Straten W., Britton M.C. \Journal{2001}{\MNRAS}{326}{358}.

Fish V., Reid M.J., Argon A.L., Menten K.M. \Journal{2003}{\ApJ}{596}{328}.

Hamilton P.A.,   Lyne A.G. \Journal{1987}{\MNRAS}{224}{1073}.

Han J.L., Beck R., Berkhuijsen E.M. \Journal{1998}{\AAp}{335}{1117}.

Han J.L., Ferriere K., Manchester R.N. (2003) {\ApJ}, submitted.

Han J.L., Manchester R.N., Lyne A.G., Qiao G.J. \Journal{2002}{\ApJ}{570}{L17}.

Han J.L., Qiao G.J. \Journal{1994}{\AAp}{288}{759}.

Han J.L., Manchester R.N., Berkhuijsen E.M., Beck R.
\Journal{1997}{\AAp}{322}{98}.

Han J.L., Manchester R.N., Qiao G.J. \Journal{1999}{\MNRAS}{306}{371}.

Han J.L., Wielebinski R. (2002) {\it Chinese Journal of A\&A} {\bf 2}, 293.

Hildebrand R.H., Davidson J.A., Dotson J.L., Dowell C.D., Novak G.,
Vaillancourt J. \Journal{2000}{\PASP}{112}{1215}.

Indrani C., Deshpande A.A. (1998) {\it New Astronomy} {\bf 4}, 33.

Kramer M., Bell J.F., Manchester R.N. et al. \Journal{2003}{\MNRAS}{342}{1299}.

Krause F., Raedler K.H. 1980, {\it Mean-field Magnetohydrodynamics and
  Dynamo Theory} (Oxford: Pergamon Press).

Kulsrud R.M. \Journal{1999}{\ARAaAp}{37}{37}.

Lyne A.G., Smith F.G. \Journal{1968}{\Nat}{218}{124}.

Lyne A.G., Smith F.G. \Journal{1989}{\MNRAS}{237}{533}.

Lyne A.G., Manchester R.N., Lorimer D.R., Bailes M., D'Amico N., Tauris
T.M., Johnston S., Bell J.F., Nicastro L. \Journal{1998}{\MNRAS}{295}{743}.

Manchester R.N. \Journal{1972}{\ApJ}{172}{43}.

Manchester R.N. \Journal{1974}{\ApJ}{188}{637}.

Manchester R.N., Lyne A.G., Camilo F. et al. \Journal{2001}{\MNRAS}{328}{17}.

Manchester R.N., Lyne A.G., D'Amico N., Bailes M., Johnston S., Lorimer
D.R., Harrison P.A., Nicastro L., Bell
J.F. \Journal{1996}{\MNRAS}{279}{1235}.

Mitra D., Wielebinski R., Kramer M., Jessner A. \Journal{2003}{\AAp}{398}{993}.

Morris D.J., Hobbs G., Lyne A.G. et al. \Journal{2002}{\MNRAS}{335}{275}.

Novak G., Chuss D.T., Renbarger T., Griffin G.S.,
 Newcomb M.G., Peterson J.B.,
 Loewenstein R.F., Pernic D., Dotson J.L. \Journal{2003}{\ApJ}{583}{L83}.

Rand R.J., Kulkarni S.R. \Journal{1989}{\ApJ}{343}{760}.

Rand R.J., Lyne A.G. \Journal{1994}{\MNRAS}{268}{497}.

Rees M. \Journal{1987}{\qjras}{28}{197}.

Reid M.J., Silverstein  E.M. \Journal{1990}{\ApJ}{361}{483R}.

Ruzmaikin A.A., Sokolov D.D., Shukurov A.M. (1988) {\it Magnetic
 Fields of Galaxies} (Dordrecht: Kluwer).

Simard-Normandin M., Kronberg P.P. \Journal{1980}{\ApJ}{242}{74}.

Sofue Y., Fujinmoto M. \Journal{1983}{\ApJ}{265}{722}.

Sofue Y., Reich W., Inoue M., Seiradakis J.H. \Journal{1987}{\PASJ}{39}{95}.

Strong A.W., Moskalenko I.V., Reimer O. \Journal{2000}{\ApJ}{537}{763}.

Thomson R.C., Nelson A.H. \Journal{1980}{\MNRAS}{191}{863}.

Vall\'ee J.P. \Journal{1991}{\ApJ}{366}{450}.

Vall\'ee J.P. \Journal{1995}{\ApJ}{454}{119}.

Vall\'ee J.P. \Journal{1996}{\AAp}{308}{433}.

Weisberg J.M., Cordes J.M., Kuan B., Devine K.E., Green J.T., Backer D.C.
 (2004) {\ApJ}, in press.

Yusef-Zadeh F., Morris M. \Journal{1987}{\ApJ}{320}{545}.

\end{document}